\documentclass{cjpsuppl}
\usepackage{graphics} 
\usepackage{graphicx}
\usepackage{epsfig} 
 \begin{document}
 \slacs{.6mm}
 \title{Transverse momentum distributions in Semi-inclusive Deep Inelastic Scattering}
 \authori{Federico A. Ceccopieri}
 \addressi{Dipartimento di Fisica, Universit\'a di Parma, Viale delle Scienze, Campus Sud, 43100, Parma, Italy}
 \authorii{}    \addressii{}
 \authoriii{}   \addressiii{}
 \authoriv{}    \addressiv{}
 \authorv{}     \addressv{}
 \authorvi{}    \addressvi{}
 \headtitle{ \ldots}
 \headauthor{}
 \lastevenhead{author: title \ldots}
 \refnum{}
 \daterec{} 
 \suppl{?}  \year{2006} \setcounter{page}{1}
 \maketitle

 \begin{abstract}
  
Within a perturbative approach to Quantum Chromodynamics (QCD),
we show how to extend ordinary DGLAP longitudinal evolution equations
to include the radiative transverse momentum generated in the collinear branching regime.
Considering Semi-inclusive Deep Inelastic Scattering as a reference process, we perform 
such a generalization both in the current and in the target fragmentation region.
These distributions are then used to predict semi-inclusive Deep Inelastic Scattering cross-sections 
onto the whole phase space of the detected hadron. 

\end{abstract}

\section{Introduction} 
Semi-inclusive Deep Inelastic Scattering (SIDIS) cross-sections, at variance with inclusive DIS,   
are differential on final state hadron variables and thus are a rich source 
of information on perturbative and non-perturbative QCD dynamics.    
In presence of a hard scale, set by the exchanged virtual boson in DIS, perturbative QCD
predicts, thank to factorization theorem \cite{fproof}, the SIDIS cross-sections.
The latter theorem garantees that non-perturbative distributions 
which parametrize soft hadronic wave functions are process independent and obey renormalization group equations, 
\textsl{i.e.} DGLAP evolution equations \cite{DGLAP}.
Furthemore final state hadrons in SIDIS are expected to have a sizeable transverse momentum
generated by QCD radiation  and this effect is indeed observed in experimental data, see for istance \cite{EMC}.
In the following we show how to extend renormalization group equations to include 
transverse degree of freedom. 
We perform such an extension for space and time-like evolution equations 
in the current framentation \cite{SIDIS_start,SIDIS_Cij},
where factorization theorem \cite{fproof} strictly applies.
In the target fragmentation region a perturbative description of the SIDIS process can be achivied 
through fracture functions \cite{Trentadue_Veneziano}, which obey an inhomogeneous evolution equations 
and for which collinear and soft factorization has been proven \cite{Fact_M_coll,Fact_M_soft}.
We thus extend fracture functions to include transverse degree of freedom and  generalize
their evolution equations. The SIDIS cross-section is then presented at leading logarithmic 
accuracy in term of transverse momentum dependent (TMD) distributions.
Numerical methods for solving the evolution equations and physical issues related 
to the obtained solutions are discussed.
 
 \section{Trasnverse momentum dependent evolution equations}
Ordinary DGLAP evolution equations \cite{DGLAP} resum leading logarithm (LL) of the type 
$[(\alpha_s/2\pi)\log (Q^2/\mu^2_{F})]^{n}$ originating
from quasi-collinear partons emission configurations, where $\mu^2_{F}$ represents the factorization scale.
Leading contributions are obtained when 
the virtualities of the partons in the ladder are strongly ordered, 
$Q^2\gg k_{n,\perp}^2 \gg \dots \gg k_{1,\perp}^2$. 
At each branching, the emitting parton thus acquires a transverse momentum 
relative to its initial direction. 
The \textsl{radiative generated} transverse momentum 
can be taken into account through transverse momentum dependent evolution equations \cite{BCM}:
\begin{eqnarray}
\label{dglap_TMD_time}
Q^2 \frac{\partial \mathcal{D}_{i}^{h}(z_h,Q^2,\bm{p_{\perp}})}{\partial Q^2}&=&
\frac{\alpha_s(Q^2)}{2\pi}\int_{z_h}^1 \frac{du}{u} 
P_{ij}(u,\alpha_s(Q^2))\cdot\\
&& \cdot \frac{d^2 \bm{q_{\perp}}}{\pi}\,\delta(\,u(1-u)Q^2-q^2_{\perp})\,
\mathcal{D}_{j}^{h}\Big(\frac{z_h}{u},Q^2,\bm{p_{\perp}}-\frac{z_h}{u} \bm{q_{\perp}} \Big).
\nonumber
\end{eqnarray}
One-particle distributions $\mathcal{D}_{i}^{h}(z_h,Q^2,\bm{p_{\perp}})$ 
in eq.(\ref{dglap_TMD_time}) give
the probability to find, at a given scale $Q^2$, a hadron $h$ 
with longitudinal momentum fraction $z_h$ and transverse momentum $\bm{p}_{\perp}$ 
relative to the parent parton $i$.
$P_{ij}(u)$ are the time-like splitting functions which, at least at LL accuracy, 
can be interpreted as the probabilities
to find a parton of type $i$ inside a parton of type $j$ and are 
expressed as a power series of the strong running coupling,
$P_{ij}(u)=\sum_{n=0}\alpha_s^{n}(Q^2)P_{ij}^{(n)}(u)$.
The order $n$ of the expansion of the splitting function matrix $P_{ij}(u)$ 
actually sets the accuracy of the evolution equations. 
The radiative transverse momentum $q_{\perp}$ at each branching 
satisfies the invariant mass constraint  $q^2_{\perp}=u\,(1-u)\,Q^2$ 
while the \textsl{transverse} arguments of   $\mathcal{D}_{i}^{h}(z_h,Q^2,\bm{p_{\perp}})$
are derived taking into account the the Lorentz boost of transverse momenta.
The unintegrated distributions fulfil the normalization:
\begin{equation}
\label{timelike_norm}
\int d^2 \bm{p}_{\perp} \mathcal{D}_{i}^{h}(z_h,Q^2,\bm{p}_{\perp})=\mathcal{D}_{i}^{h}(z_h,Q^2)\,.
\end{equation}
This property garantees that we can recover ordinary integrated distributions from 
unintegrated ones. The opposite statement is not valid since eq.(\ref{dglap_TMD_time}) 
contains new physical information. 
In analogy to the time-like case we consider now a initial state parton $p$
in a incoming proton $P$ which undergoes a hard collision, 
the reference frame being aligned along the incoming proton axis.
We thus generalize eq.(\ref{dglap_TMD_time}) to the space-like case \cite{our_work}:
\begin{eqnarray}
\label{dglap_TMD_space}
Q^2 \frac{\partial \mathcal{F}_{P}^{i}(x_B,Q^2,\bm{k_{\perp}})}{\partial Q^2}
&=&\frac{\alpha_s(Q^2)}{2\pi}\int_{x_B}^1 \frac{du}{u^3} 
P_{ji}(u,\alpha_s(Q^2))\cdot\nonumber\\
&&\cdot \frac{d^2 \bm{q_{\perp}}}{\pi}\,\delta(\,(1-u)Q^2-q^2_{\perp})
\,\mathcal{F}_{P}^{j}\Big(\frac{x_B}{u},Q^2, \frac{\bm{k}_{\perp}-\bm{q}_{\perp}}{u} \Big)\,.
\end{eqnarray}
One-particle distributions $ \mathcal{F}_{P}^{i}(x_B,Q^2,\bm{k_{\perp}})$ 
in eq.(\ref{dglap_TMD_space}) give
the probability to find, at a given scale $Q^2$, a parton  $i$ 
with longitudinal momentum fraction $x_B$ and transverse momentum $\bm{k}_{\perp}$ 
relative to the parent hadron.
The unintegrated distributions fulfil a condition
analogous to the one in eq.(\ref{timelike_norm}), \textsl{i.e.} :
\begin{equation}
\label{spacelike_norm}
\int d^2 \bm{k}_{\perp} \mathcal{F}_{P}^{i}(x_B,Q^2,\bm{k}_{\perp})=\mathcal{F}_{P}^{i}(x_B,Q^2)\,.
\end{equation}
We note that the inclusion of transverse momentum does not affect longitudinal 
degrees of freedom since partons  
always degrade their fractional momenta in the perturbative branching processes.

The perturbative approach can also be extended in the target fragmentation region
introducing  new non-perturbative distributions \cite{Trentadue_Veneziano},
\textsl{i.e.} fracture functions $\mathcal{M}^{i}_{P,h}(x,z,Q^2)$,
whose collinear and soft factorization has been proven respectively in \cite{Fact_M_coll,Fact_M_soft}\,.
These functions express the conditional probability 
of finding, at a scale $Q^2$, a parton $i$ with momentum fraction $x$ in a proton $P$
while a hadron $h$ with momentum fraction $z$ is detected. 
Fracture functions are shown to obey inhomogeneous evolution equations \cite{Trentadue_Veneziano}.  
We generalize these distributions to contain also transverse degrees of freedom. The fracture functions 
$\mathcal{M}^{i}_{P,h} (x,\bm{k}_{\perp},z,\bm{p}_{\perp},Q^2)$ give the conditional probability 
to find in a proton $P$, at a scale $Q^2$, a parton with momentum fraction $x$ and transverse momentum 
$\bm{k}_{\perp}$ while a hadron $h$, with momentum fraction $z$ 
and transverse momentum $\bm{p}_{\perp}$, is detected. 
Under these assumptions the following evolution equations can thus be derived \cite{our_work}:
\begin{eqnarray}
\label{M-evo_long+tra}
&&Q^2\frac{\partial \mathcal{M}^{i}_{P,h}
(x,\bm{k}_{\perp},z,\bm{p}_{\perp},Q^2)}{\partial Q^2}=
\frac{\alpha_s(Q^2)}{2\pi} \Bigg\{ \int_{\frac{x}{1-z}}^{1} \frac{du}{u^3} \,P_{ji}(u)\, 
\int\frac{d^2 \bm{q}_{\perp}}{\pi}\, \delta(\,(1-u)Q^2-q_{\perp}^2)\cdot\nonumber\\
&&\cdot\mathcal{M}^{j}_{P,h}\Big(\frac{x}{u},\frac{\bm{k}_{\perp}-\bm{q}_{\perp}}{u},
z,\bm{p}_{\perp},Q^2 \Big)+ \int_{x}^{\frac{x}{x+z}} \frac{du}{x(1-u)u^2} \hat{P}^{i,l}_{j}(u)
\frac{d^2 \bm{q}_{\perp}}{\pi}\,\delta(\,(1-u)Q^2-q_{\perp}^2) \cdot \nonumber\\
&& \cdot \mathcal{F}_{P}^{j} 
\Big(\frac{x}{u},Q^2,\frac{\bm{k}_{\perp}-\bm{q}_{\perp}}{u}, \Big)\,
\mathcal{D}_{l}^{h}\Big(\frac{zu}{x(1-u)},Q^2,\bm{p}_{\perp}-\frac{zu}{x(1-u)}\,
\bm{q}_{\perp} \Big)\Bigg\} \,. 
\end{eqnarray}  
The homogeneous term has a pure non-perturbative 
nature since involves the fragmentation of the proton remnants into the hadron $h$, 
while the active parton $i$ evolve towards the hard boson vertex. 
The inhomogeneous one takes into account the production of the hadron $h$ from a time-like cascade of 
parton $j$. In this latter term $\hat{P}^{i,l}_{j}(u)$
stand for the unsubtracted splitting functions \cite{unregAP}\,.
Fracture functions fulfil the normalization condition:
\begin{equation}
\label{M_norm}
\int d^2 \bm{k}_{\perp}\int d^2 \bm{p}_{\perp} \mathcal{M}^{i}_{P,h}
(x,\bm{k}_{\perp},z,\bm{p}_{\perp},Q^2) =  \mathcal{M}^{i}_{P,h}
(x,z,Q^2) \,,
\end{equation}
as direct consequence of the kinematics of both terms in the evolution equations, eq.(\ref{M-evo_long+tra}).
To be definite we must say that the soft and collinear factorization of these objects 
is at the moment only a \textsl{conjecture}. 

\newpage
\section{Solutions to evolution equations and phenomenology}

In our opinion the study of target fragmentation phenomenology is indeed the most intruiguing aspects 
offered by the novel equations we have proposed. However, due to lack of the aforementioned factorization 
theorem for TMD fracture functions,  we think that current fragmentation phenomenology 
is a good benchmark for our evolution equations since a great amount of data to confront with is available. 
In what follows we thus describe the method for solving the evolution equations in the space-like case only,
eq.(\ref{dglap_TMD_space}), since similar arguments applie also in the time-like case, eq.(\ref{dglap_TMD_time}). 
Solutions in both kinematics allows us to compare directly with current fragmentation phenomenology.
   
As in the longitudinal case, distributions at a
scale $Q^2>Q_0^2$ are calculable if we provide a non-perturbative input density at some arbitrary scale $Q_0^2$.
We assume as initial condition a longitudinal distribution functions \cite{GRV,Kretzer} 
times a $x$-independent gaussian transverse factor:  
\begin{equation}
\label{ansatz}
\mathcal{F}_{P}^{i}(x_B,Q^2,\bm{k}_{\perp})
=  \mathcal{F}_{P}^{i}(x_B,Q^2)
\;\frac{\makebox{e}^{-\bm{k}_{\perp}^2/<\bm{k}_{\perp,i}^2>}}{\pi<\bm{k}_{\perp,i}^2>} \,.
\end{equation}
The gaussian $\bm{k}_{\perp}$-distribution
is used to model partons intrinsic momenta inside hadrons at the initial scale. This assumption is reasonable 
in the valence region $x\in\mathcal{O}(1)$ but questionable in the gluon-dominated small-$x$ range, 
a relevant issue for HERA kinematics, 
where the transverse factor, we argue, should become in some way $x$-dependent and also modified in the shape.  
Average transverse momentum is then set to $<\bm{k}_{\perp, q_{i},g}^2> \simeq 0.25$ $GeV^2$.
We choose a finite difference method for solving the evolution
equations since it give us a direct control on the starting distributions and we set the splitting functions 
$P_{ij}^{(n)}(x)$ at lowest order ($n=0$) in their expansion.
The r.h.s. of the evolution equations are thus integrated on a 
bidimensional grid in $(x, \bm{k}_{\perp}^2)$. At each iteration on the $Q^2$ variable, 
the normalization conditions, eq.(\ref{spacelike_norm}),
is checked to reproduce ordinary longitudinal distributions. 
We solve the $(2n_f+1)-$dimensional matrix equations in the space of 
quarks, antiquarks and gluons. 
The resulting space-like TMD distributions  show a $x-$dependent broadening 
of the transverse momentum spectrum with increasing $Q^2$. Furthermore the factorized  
form of eq.(\ref{ansatz}) is not preserved under evolution. This result is expected since 
the transverse part of the TMD distributions, eq.(\ref{dglap_TMD_space}), mixes longitudinal and transverse variables.   
The spectrum tail shows a inverse-powerlike behaviour 
of the type $(\bm{k}_{\perp}^2)^{-\alpha}$, with $1<\alpha<2$ depending on x, at high $\bm{k}_{\perp}^2$. 
Noticeable deviations from a guassian behaviour at incresing $Q^2$ are observed. 
The more pronunciated broadening in the transverse spectrum is more appreciable  
in the TMD gluon distribution at small $x$. 
Such an effect is expected since $P_{gq}(x),P_{gg}(x) \simeq 1/x$ as $x\rightarrow0$.

As previosly stated, with TMD distributions at hand, we can directly reconstruct
the structure functions $H_{i=1,..,4}(x_B,z_h,Q^2,\bm{P}_{h\perp})$ \cite{h2_mulders}, 
in the current fragmentation region according to 

\newpage
\begin{eqnarray} 
\label{kt_sidis_fact}
&&H_2(x_B,z_h,\bm{P}_{h\perp},Q^2)=\sum_{i=q,\,\bar{q}}
e_q^2\int d^2\bm{k}_{\perp} d^2\bm{p}_{\perp} \delta^{(2)} 
(z_h \bm{k}_{\perp}+\bm{p}_{\perp} - \bm{P}_{h\perp})\cdot\nonumber\\
&&\quad\quad\quad\quad\quad\quad\quad\quad\quad
\cdot\;\mathcal{F}_{P}^{i}(x_B,\mu^2_F,\bm{k}_{\perp},) \; \mathcal{D}^{h}_{i}(z_h,\mu^2_D,\bm{p}_{\perp}) 
\; C(Q^2,\mu^2_F,\mu^2_D) \,,
\end{eqnarray}    
where the factorization of eq.(\ref{kt_sidis_fact}) in terms of TMD distributions has been 
explictely proven in \cite{Ji} and is accurate up to powers in $(P_{h\perp}^2/Q^2)^n$ for  
transverse momenta $|P_{h\perp}|\simeq\Lambda_{QCD}$. 
The standard SIDIS variables are defined as $x_B=Q^{2}/(2 P \cdot q)$
and $z_h=(P\cdot P_h) / (P\cdot q)$ . Since we are using LL evolution 
equations, we set $C=1$ and the factorization scales $\mu^2_{F}=\mu^2_{D}=Q^2$.
In our opinion is interesting to note that gluon, not directly coupled to the lepton
current at LL in eq.(\ref{kt_sidis_fact}), partecipate
the process through the evolution equations, eq.(\ref{dglap_TMD_space}) and eq.(\ref{dglap_TMD_time}).
We are thus, even at lowest order, indirectly sensitive to the transverse spectra of the gluon
which manifests its dynamical effects mostly on sea-quarks transverse spectra at small $x$.
In the opposite kinematic region, $x\rightarrow1$, we are instead sensitive to soft gluon resummation
effects which can effectively taken into account in our evolution equations setting 
$\alpha_s(Q^2)\rightarrow\alpha_s((1-x)Q^2)$\, \cite{CT}.
Once resummation effects are taken into account, the valence region it is the best suited kinematical 
region in which extract the non-perturbative quark intrinsic transverse momentum.
  
Eq. (\ref{kt_sidis_fact}) can be shown to include also target fragmentated hadrons using TMD fracture functions
\cite{our_work}. In this case, as previously stated, a factorization theorem for TMD fracture functions 
is lacking and the factorization is only \textsl{conjecturized}.

 \section{Conclusions}

We have shown that radiative transverse momentum can be quantitatively taken 
into account  using TMD DGLAP evolution equations and that the resulting 
TMD distributions show a  $\bm{k}_{\perp}^{2}-$smearing due to pQCD dynamics.
The extension can be performed both in the current and in the target fragmentation region. 
Delicate issue concerning the small-$x$ gluon distribution, impact of soft gluon resummation 
to gauge properly intrinsic transverse momentum of partons and peculiar feature of the transverse
spectrum of TMD distributions are presented.
The formalism we have developed can be extended to the polarized
case both in current and in the target region\cite{sassot_polarized}.
Furthemore, if factorization holds for TMD fracture functions,  we will have a multi-dimensional
description of SIDIS cross-sections, valid onto the whole phase space of the produced hadron.
A detailed derivation of our results and a complete phenomenological study will appear in a 
forthcoming paper \cite{newpaper} .

 \bigskip

 {\small I would like to thank Prof. Miroslav Finger for having invited me at the Conference. 
I would like also to thank Prof. Luca Trentadue for discussions on such topics.}

 \end{document}